\documentclass[aps,prl,twocolumn,floatfix,superscriptaddress,showpacs]{revtex4-1}

\usepackage{graphicx}
\usepackage[centertags]{amsmath}
\usepackage{amsfonts}
\usepackage{amssymb}
\usepackage{bm}
\usepackage{dcolumn}
\usepackage{epstopdf}

\newcommand{\aveI}{\langle I \rangle}
\renewcommand{\vec}[1]{\boldsymbol{#1}}

\begin{document}

\title{Backscattering Between Helical Edge States via Dynamic Nuclear
Polarization}

\author{Adrian \surname{Del Maestro}}
\affiliation{Department of Physics, University of Vermont, Burlington, VT 05405,
USA}

\author{Timo Hyart}
\affiliation{Institut f\"ur Theoretische Physik, Universit\"at Leipzig, D-04103,
Leipzig, Germany}

\author{Bernd Rosenow}
\affiliation{Institut f\"ur Theoretische Physik, Universit\"at Leipzig, D-04103,
Leipzig, Germany}

\date{\today}

\begin{abstract}
We show that that the non-equilibrium spin polarization of one dimensional
helical edge states at the boundary of a two dimensional topological insulator
can dynamically induce a polarization of nuclei via the hyperfine interaction.
When combined with a spatially inhomogeneous Rashba coupling, the steady state
polarization of the nuclei produces backscattering between the topologically
protected edge states leading to a reduction in the conductance which persists
to zero temperature.  We study these effects in both short and long edges,
uncovering deviations from Ohmic transport at finite temperature and a current noise
spectrum which may hold the fingerprints for experimental verification of the
backscattering mechanism.
\end{abstract}


\maketitle

% ===============================================================================
% Introduction
% ===============================================================================

Topological insulators have a bulk band gap, but cannot be adiabatically
connected to the conventional insulators without closing the energy gap
\cite{Kane05, Bernevig:2006ui, KaneRMP, ZhangRMP} because they are
characterized by a non-trivial topological number. The quantum spin Hall (QSH)
insulators belong to the class of two dimensional time reversal (TR) invariant
$\mathbb{Z}_2$ topological insulators. They support a pair of gapless helical
edge states, which give rise to unique, experimentally confirmed
\cite{Konig:2007hs,Konig:2008bz, Roth:2009bg, Brune:2012fi} electrical
properties.  Due to their helical nature, QSH edge states are protected against
elastic back-scattering from TR symmetric perturbations: there is destructive
interference between two time reversed amplitudes, one where spin is rotated in
a, say, clockwise manner and a second one with anti-clockwise rotation. In the
interference term, this amounts to a full rotation of a spin-$1/2$,
corresponding to a factor of minus one, which cancels the contribution of the
direct term. Backscattering due to TR breaking perturbations
\cite{MaciejkoPRB10,Delplace:2012ww,Tanaka11,Hattori} and from inelastic
processes
\cite{Roth:2009bg,Jiang:2009fr,Budich:2012dg,Maciejko:2009kw,Crepin:2012vj,Lezmy12,Schmidt:2012iw}
is however allowed.

The protection of the edge states in the presence of TR symmetry can be
exemplified by considering the scattering due to a spatially random Rashba
spin-orbit coupling \cite{Sherman:2003hx,Golub:2004il,Strom:2010to,Rothe:2010cw}. 
In Fourier space, the matrix element for this process
includes a factor $(k + k^\prime)$, where $k$ and $k^\prime$ are the momenta of
initial and final states, respectively. For a TR invariant Hamiltonian, the
dispersion relation satisfies  $\epsilon_k = \epsilon_{-k}$, such that energy
conservation enforces $k^\prime = -k$ implying a vanishing matrix element for
backscattering. In the presence of an in-plane magnetic field however, there is 
a Zeeman splitting  $2 \Delta =  g \mu_\text{B} B$
between opposite spin directions. For a linearized dispersion $\epsilon(k) =
\hbar v k$ this gives rise to a non-vanishing prefactor $(k + k^\prime) = 2
\Delta / \hbar v$ and yields a finite mean free path $\ell \propto 1/\Delta^2$.
This mechanism can explain the experimentally observed dependence of the
quantum spin Hall edge conductance on an in-plane magnetic field
\cite{Konig:2007hs,Konig:2008bz}.

\begin{figure}[t]
\begin{center}
\includegraphics[width=\columnwidth]{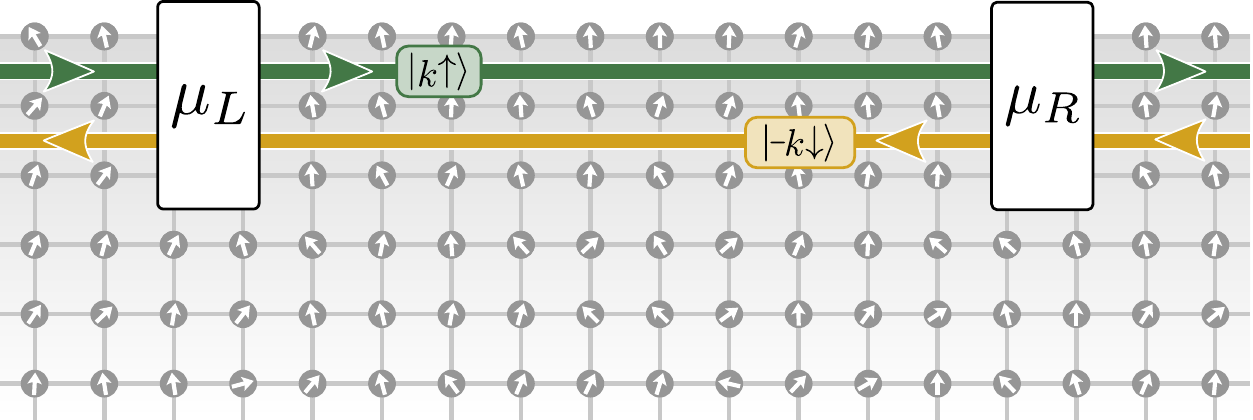}
\caption{\label{fig:edgeStates} Quantum spin Hall edge states dynamically
polarize nuclear spins via a hyperfine interaction.  Although the edge states
are drawn as sharp lines, their wavefunctions may overlap a large number of
nuclei.  Only one edge is shown, but there is an equivalent and independent
transport channel at the opposite edge, where the spin polarizations are
reversed.}
\end{center}
\end{figure}

In this letter, we discuss the influence of dynamically polarized nuclear spins
on the conductance of QSH edge channels depicted in Fig.~\ref{fig:edgeStates}.
%
% ----------------------------------------------------------------------------
% ----------------------------------------------------------------------------
%
We argue that flip-flop scattering between electronic and nuclear spins creates
a dynamic nuclear polarization, which has the same effect as an external Zeeman
field and gives rise to backscattering between helical edge states. The
magnitude of the nuclear spin polarization is determined by the ratio of bias
voltage and temperature. Thus, for a long edge of length $L \gg \ell$, the local
temperature profile governed by a balance between Joule heating and electronic
heat transport determines the conductance. We find a non-linear current-voltage
characteristic at finite temperature and a peculiar relation between applied
voltage and noise power, which together with hysteresis in the current-voltage
characteristic are signatures of backscattering due to a dynamic nuclear
polarization.  Importantly, this mechanism stays effective in the
low-temperature limit, similar to spin-flip scattering in $\nu=2$ quantum Hall
edges \cite{Wald:1994bz}.

% ===============================================================================
% Model
% ===============================================================================
At low energies in the presence of a nuclear spin polarization, the helical edge
electrons have an effective Dirac spectrum governed by
\begin{equation}
H_0 = \int d x\, \psi^{\dag}_{\alpha}(x) \sigma^z_{\alpha \beta}  \left(-iv \hbar
\partial_x + \frac{1}{2}  A \aveI\right) \psi^{\phantom{\dag}}_{\beta}(x), \label{eq:H0}
\end{equation}
where $\psi^{\dag}_{\alpha = \uparrow,\downarrow}(x)$ is a fermionic field
corresponding to a right moving (spin-up) or left moving (spin-down) electron
with Fermi velocity $v$, $\sigma^z$ is a Pauli matrix in spin space,  and
repeated indices are summed over.  The polarization of nuclear spins is
generated via the hyperfine dynamic Hamiltonian \cite{Slichter:1990pr}
\begin{equation}
H_{\text{flip}} = \frac{A}{2 \rho_n} \sum_{i}
\psi^\dag_{\alpha}(x_i) \left( \sigma^{+}_{\alpha \beta} I^{-}_{i}
+ \sigma^{-}_{\alpha \beta} I^{+}_{i} \right)
\psi^{\phantom{\dag}}_\beta(x_i)
\label{eq:Hflip}
\end{equation}
where $\vec{I}_{_i}$ is the spin operator for a nucleus located at position
$\vec{r}_i = (x_i,y_i,z_i)$, $\sigma^\pm = (\sigma_x \pm i \sigma_y)/2$,  and
$\rho_n$ is the effective one dimensional density of nuclei seen by the edge
electrons.  Spin flips are generated by this effective contact term, where the
strength of the hyperfine interaction, $A$, can be measured using optical or
electron paramagnetic resonance experiments \cite{Stone:2005tn}.
Backscattering between right movers and left movers can be induced by a
spatially dependent Rashba spin-orbit coupling term that can arise as a result
of electrostatic potential disorder
\cite{Sherman:2003hx,Golub:2004il,Strom:2010to,Rothe:2010cw}.  The interactions
are described by
\begin{equation} H_{R} = \int d x\, \psi^{\dag}_{\alpha}(x)
\sigma^{y}_{\alpha\beta} \{ a(x), i\partial_x \}
\psi^{\phantom{\dag}}_{\beta}(x) \label{eq:HRashba} \end{equation}
where the strength of the inhomogeneous coupling is set by the scale of its
local correlations: $\langle a(x) a(x')\rangle = V_R \delta(x-x')$.  The total
Hamiltonian is given by $H = H_0 + H_{\text{flip}} + H_{\text{R}}$ where the
effective Zeeman gap originating from the polarized nuclei $A \aveI$ will need
to be determined from the steady state of nuclear polarization.

% ===============================================================================
% Spin flip rate and nuclear polarization
% ===============================================================================

To this end, we begin by calculating the net scattering rate per unit length,
$\Gamma$, describing the transfer of electrons between the right and left
moving spin polarized edge states, $|k\!\uparrow \rangle$ and $|\!-k^\prime\!\downarrow
\rangle$, due to the interaction between the electron and the nuclear spins
$H_{\textrm{flip}}$.  Employing Fermi's golden rule we obtain
\cite{Overhauser:1953cu,Tripathi:2008br,Lunde:2012he}
\begin{eqnarray}
\Gamma&=&\frac{A^2}{8 \pi \hbar^3 v^2 \rho_n}  \int d \epsilon \ \bigg\{
    \left(1/2 - \langle I(t) \rangle \right) f_+ (\epsilon) [1-f_-(\epsilon)]
    \nonumber\\ && \hspace{1cm} - \left(1/2 + \langle I(t) \rangle \right)  f_-
    (\epsilon) [1-f_+(\epsilon)] \bigg\},
\end{eqnarray}
where $f_\pm(\epsilon)$ are the distribution functions for the right and left
moving edge states, respectively.

Because each scattering event is associated with a nuclear spin flip, the
time evolution of the nuclear spin polarization can be described with a rate
equation
$
\partial_t \langle I(t) \rangle=\Gamma/ \rho_n -\Gamma_d \langle I(t)
\rangle/\rho_n ,
$
where $\Gamma_d$ is a phenomenological nuclear spin relaxation rate
encompassing all other mechanisms that lead to nuclear spin flips.
In quantum Hall systems, the nuclear spin relaxation rate is
typically much smaller than the spin flip scattering rate that originates from
the dipole interaction between the nuclear spins and the spin polarized edge
states \cite{Wald:1994bz}.  Due to their similarity with the QSH edge states
considered here, we expect that $\Gamma_d \ll \Gamma$.

The rate equation for the nuclear spin polarization can be solved for
arbitrary electron distribution functions $f_\pm(\epsilon)$. Assuming that the
nuclear spin polarization is initially zero, the time evolution is described by
$\langle I(t) \rangle = \langle I \rangle(1-e^{-t/\tau_n})$, where
\begin{eqnarray}
\langle I \rangle &=& \frac{\int d \epsilon
[f_+ (\epsilon)-f_-(\epsilon)]/2}{\int d\epsilon \{f_+ (\epsilon) +f_- (\epsilon)
[1 -2 f_+(\epsilon)]\} + \frac{\Gamma_d 8 \pi \hbar^3 v^2\rho_n }{A^2}}, \nonumber \\
\tau_n^{-1}&=&\frac{A^2 \int d \epsilon \  \{f_+ (\epsilon) +
f_- (\epsilon)[1 -2 f_+(\epsilon)]\}  }{8 \pi \hbar^3 v^2 \rho_n^2}+
\frac{\Gamma_d}{\rho_n}. \nonumber
\end{eqnarray}
It is difficult to obtain a reliable estimate for the characteristic time-scale
$\tau_n$, where the steady-state nuclear polarization $\langle I
\rangle$ is achieved. Based on the experiments in the quantum Hall
systems \cite{Wald:1994bz}, we expect that the first term in $\tau_n^{-1}$
dominates and $\tau_n$ is on the order of tens or hundreds of seconds.

% ===============================================================================
% Nuclear spin polarization for short edge
% ===============================================================================

To evaluate the effect of a finite nuclear polarization $\langle I \rangle$ on electrical transport, 
we first concentrate on the case of a short edge, where the distance between contacts
$L$ is much shorter than the mean free path associated with the spin flip
scattering. The distribution functions are then determined by the Fermi
functions $f_\pm(\epsilon)=f_0(\epsilon- \mu_{L/R}, T)$, where $f_0(\epsilon,
T)=[1+\exp(\epsilon/k_\text{B} T)]^{-1}$, and $\mu_{L/R}$ are the chemical potentials
of the left/right reservoirs, respectively. At zero temperature $\langle I
\rangle=\frac{1}{2}[1+\frac{\Gamma_d 8 \pi \hbar^3 v^2 \rho_n}{A^2
(\mu_L-\mu_R)}]^{-1}$, so that in the absence of relaxation, the nuclear
spins become fully polarized. For $\Gamma_d=0$, the temperature dependence of
the nuclear spin polarization is described by  $\langle I
\rangle=\frac{1}{2} \tanh\big[(\mu_L-\mu_R)/2k_\text{B}T \big]$.

\begin{figure}[t]
\begin{center}
\includegraphics[width=\columnwidth]{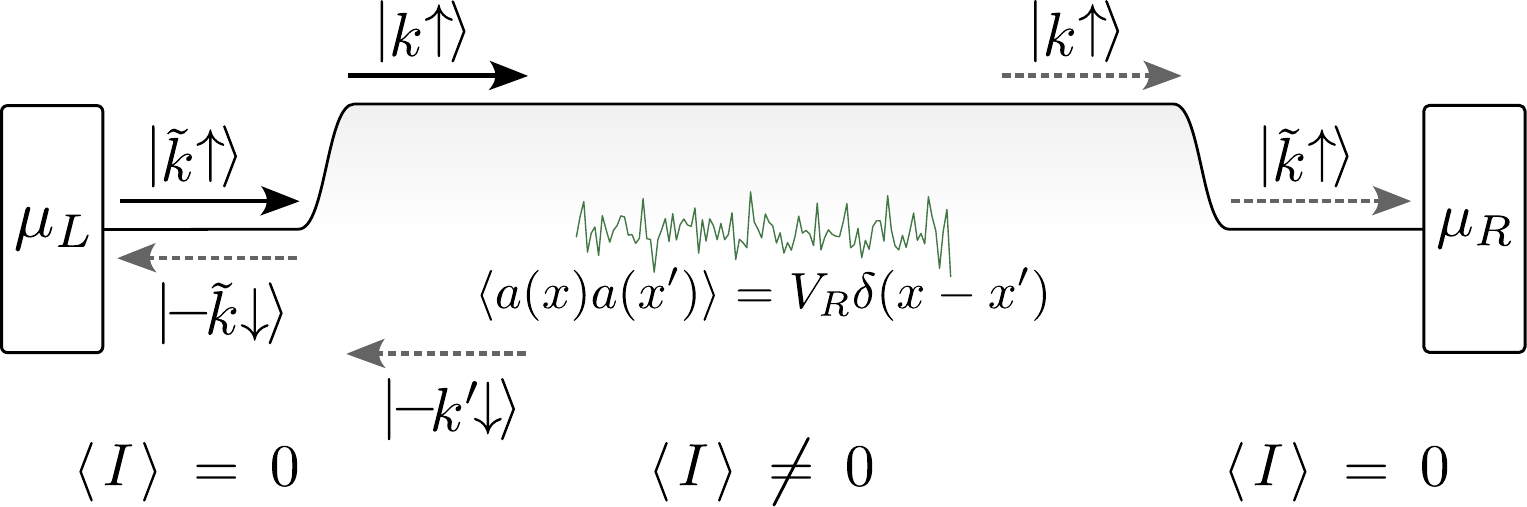}
\caption{ \label{fig:scatteringStates} An outline of sequential scattering
processes for the helical edge states.  It is assumed that the nuclei are
unpolarized $(\langle I \rangle = 0)$ in the leads but are dynamically
polarized $(\langle I \rangle \ne 0)$ between them in a region with potential
disorder giving rise to random Rashba scattering.}
\end{center}
\end{figure}

% ===============================================================================
% The Short Wire
% ===============================================================================
We now turn to evaluating the backscattering rate due to the random Rashba spin
orbit coupling in the presence of a dynamically generated nuclear spin
polarization $\langle I \rangle$. Denoting initial and final electronic states
by their momenta $|k\!\uparrow \rangle$ and $|k^\prime\!\downarrow \rangle$, we
find the matrix element $\langle k^\prime | H_R | k \rangle = - i (k +
k^\prime) \, \hat{a}(k^\prime - k)/L$.  Taking into account that due to energy
conservation   $(k + k^\prime) = -   A \langle I \rangle / v \hbar$, we find
for the disorder averaged scattering rate
$W_{k k^\prime} = 2 \pi   V_R  A^2 \langle I \rangle^2
\delta(\epsilon_k - \epsilon_{k^\prime})/( \hbar^3 v^2 L)$.
The total Rashba scattering rate is then
%
%********************  total Rashba scattering rate  ********************
\begin{equation}
\frac{1}{\tau_R}\equiv \sum_{k^\prime} W_{k k^\prime} =
A^2 \langle I \rangle^2 { V_R \over \hbar^4 v^3} \ \ .
\label{rashbascattering.eq}
\end{equation}
%**************************************************************************
%
The edge conductivity due to Rashba scattering is $\sigma = (e^2 /\pi \hbar) \ell$
with $\ell = v \tau_R/2$ (the extra factor $1/2$ accounts for the difference
between single particle and transport relaxation time). It is related to the two-terminal conductance via
%
%********************  2-terminal conductance *************
\begin{equation}
G = \left( {h \over e^2} + {L \over \sigma} \right)^{-1} \ \ .
\end{equation}
%*************************************************************
%
In the limit of a short edge with $L \ll \ell$, the conductance is changed by
%
%***********************************  conductance change  ***********
\begin{equation}
\delta G = - {e^2 \over h} \, A^2  \langle I \rangle^2 { V_R \over \hbar^4 v^4} L \ \ .
\label{deltaG.eq}
\end{equation}
%*************************************************************************
%
This equation can be directly verified by calculating the change of current caused
by scattering between the spin polarized edge states.  Here, we assume
that the phase space for scattering is not changed by the presence of a spin
splitting due to the nuclear polarization.  This can be understood by
calculating scattering states from the time independent Schr\"{o}edinger
equation describing energy eigenstates as pictured schematically in
Fig.~\ref{fig:scatteringStates}.
%
% ----------------------------------------------------------------------------
% ----------------------------------------------------------------------------
%
We assume that there is no spin splitting in the reservoirs, and that it is
switched on and off in regions of the edge where there is no Rashba scattering.
Since the nuclear magnetization only couples to the electron spin, which is
assumed to be a good quantum number, a finite $\langle I \rangle$ leads to
a change in wave vector, such that Rashba scattering has a finite matrix
element. After the scattering, $\langle I \rangle$ is switched off again before
entering the reservoir, so the initial and final momentum are equal in
magnitude.

% ===============================================================================
% The Long Wire
% ===============================================================================

For long edges with $L \gg \ell$, the distribution functions of the electrons
are no longer determined by the reservoir distributions but have to be
calculated from a solution of a kinetic equation. In the following, we assume
that the Coulomb interaction between counter-propagating edge states gives 
rise to an inelastic mean free path approximately equal to the elastic one 
\cite{Bagrets:2009hc}.  Thus, it seems reasonable to assume that a long
edge region can be described locally by equilibrium distributions with position
dependent chemical potential $\mu(x)$, temperature $T(x)$,  and nuclear
polarization $\langle I \rangle (x)$. Neglecting the relaxation of nuclear
spins, the nuclear polarization is related to the difference in chemical
potential between left and right movers, $\delta \mu$, which in turn is
determined by the electrical current $j$ via $ \delta \mu = j h /e$. As current
is conserved along the edge, we have
%
%************************** long edge nuclear polarization  ******************
\begin{equation}
\langle I \rangle(x)  \ = \ {1 \over 2}
\tanh \left( {j \pi \hbar \over e  k_\text{B} T(x)} \right) \ \ .
\label{polarization_long.eq}
\end{equation}
%*******************************************************************************
%

Expressing the electron distribution function as
%
%******************  electron distribution  **************
\begin{equation}
f(k,x) = f_0(\xi_k,T(x)) + \delta f(k,x) \ \ ,
\label{distribution_long.eq}
\end{equation}
%********************************************************
%
with $\xi_k = \epsilon_k + {\rm sign}(k) A \langle I \rangle(x) - \mu(x)$
subject to the linearized steady state Boltzmann equation $ v_k \partial_x f_0
= - \delta f 2/\tau_R$, we find the solution
%
%********************  delta f *************************
\begin{equation}
\delta f(k,x) \ = \ {\tau_R \over 2} \delta(\xi_k ) \left[ \text{sign}(k) A
\partial_x \langle I \rangle -\partial_x \mu - {\xi_k \over T^2} \partial_x T
\right]  \  .
\label{deltaf.eq}
\end{equation}
%************************************************
%
The current is obtained as $j = e \int {d k \over 2 \pi} v_k \delta f$, giving
rise to a position dependent conductivity
%
%*********** conductivity ***********
\begin{equation}
\sigma (x) \ = \  {e^2 \over \pi \hbar} {\hbar^4 v^4 \over 2 V_R} {1 \over A^2
\langle I \rangle ^2(x)} \ \ .
\label{sigma_long.eq}
\end{equation}
%*************************************
%
The conductivity depends on temperature via the expectation value of the nuclear
polarization in Eq.~(\ref{polarization_long.eq}).  In principle, interaction effects 
give rise to an additional, explicit temperature dependence of the backscattering rate, 
which is not included in the following. 
In order to
determine the conductance, knowledge of the temperature profile $T(x)$ along the
edge is needed. In the limit of small bias, the edge temperature is given by the
bath temperature, $T(x) \equiv T$. Linearizing Eq.~(\ref{polarization_long.eq})
we obtain
%
%******************  small bias conductance  ************
\begin{equation}
j^3 = {e \over \pi \hbar} {2 \hbar^4 v^4 \over V_R}
\left( { e k_\text{B} T \over A \pi \hbar}\right)^2
\left( - {\partial \mu \over \partial x} \right)\ \ ,
\end{equation}
%**********************************************************
%
a nonlinear current voltage relation, where in the limit $\sigma /L \ll e^2/h$
one finds $j \propto T^{2/3}  \mu^{1/3}$.

\begin{figure}[t]
\begin{center}
\includegraphics[width=\columnwidth]{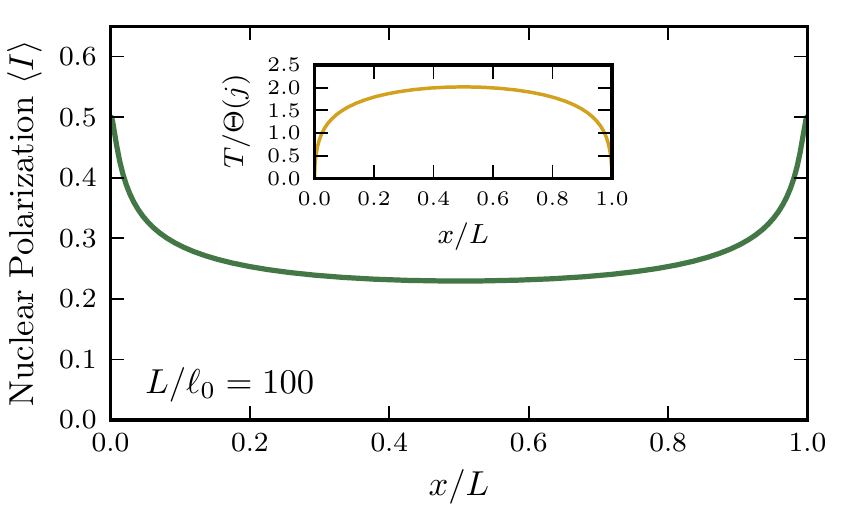}
\caption{\label{fig:magnetization}
The induced dynamic nuclear polarization $\langle I \rangle (x)$ computed from
Eq.~(\ref{polarization_long.eq}) for a long $(L/\ell_0 \gg 1)$ edge using the
dimensionless temperature profile shown in the inset.}
\end{center}
\end{figure}

In general, the temperature distribution $T(x)$ must be calculated
self-consistently, taking into account Joule heating due to the transport
current \cite{Steinbach:1996vq,vonOppen:1997vc}. Denoting the heat current by $j_Q$, it is
related to the local temperature gradient by the heat conductivity $\kappa =
\sigma T (k_\text{B}/e)^2 (\pi^2/3)$ as $j_Q =  - \kappa \partial_x T$. Joule heating
determines the divergence of the heat current according to $\partial_x j_Q =
j^2/\sigma$, such that
%
%**************************  heat current differential equation  *********************
\begin{equation}
  -    \sigma(x) \partial_x \left[ \sigma(x) \partial_x  T^2(x) \right]
  \ = \  j^2  \left( {e \over k_\text{B}}\right)^2 {6  \over \pi^2} \ \ .
  \label{Tdiff.eq}
\end{equation}
%******************************************************************
%
Together with Eqs.~(\ref{polarization_long.eq}) and (\ref{sigma_long.eq}) this
constitutes a nonlinear differential equation for the temperature profile. We
now discuss the solution of Eq.~(\ref{Tdiff.eq}) in the limit where the bath
temperature can be neglected as compared to the temperature generated by Joule
heating, e.g.~we use the boundary conditions $T(0) = T(L)=0$. Introducing the
length scale $\ell_0 = (\hbar v)^4/(2 V_R A^2)$, we define a dimensionless
coordinate $\overline{x} \equiv x /\ell_0$ and temperature variable
$t(\overline{x}) \equiv  T(x) / \Theta(j)$ with $\Theta(j) = j \pi \hbar / e
k_\text{B}$.  Eq.~(\ref{Tdiff.eq}) may now be numerically integrated yielding
the dimensionless temperature profile and nuclear polarization shown in
Fig.~\ref{fig:magnetization} for fixed $L/\ell_0 \gg 1$.
%
% ----------------------------------------------------------------------------
% ----------------------------------------------------------------------------
%

\begin{figure}[t]
\begin{center}
\includegraphics[width=\columnwidth]{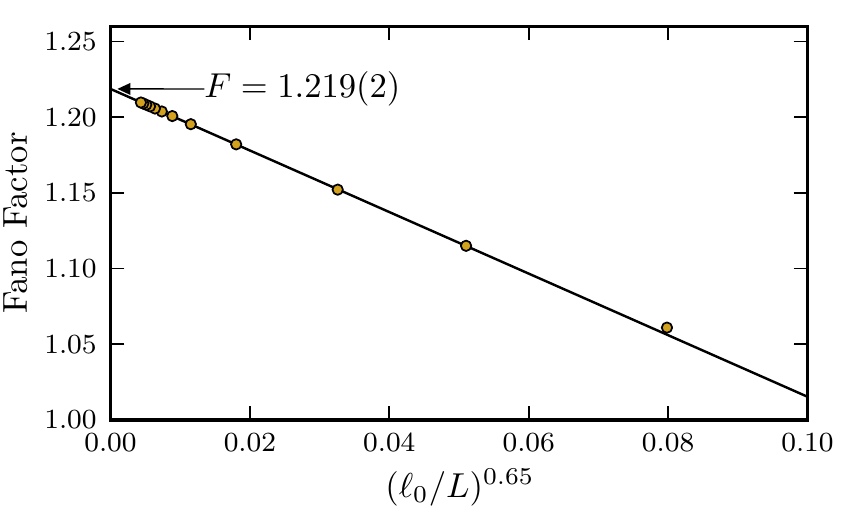}
\caption{\label{fig:fano} The Fano factor calculated from
Eq.~(\ref{eq:currentNoise}) is found to be parametrized by the finite size
scaling form $F = 1.219(2) - 2.03(3) (\ell_0/L)^{0.65(3)}$.}
\end{center}
\end{figure}

The chemical potential difference $\Delta \mu$ between the left and right
reservoirs is related to the current via
%
%************************  chemical potential vs. current *********
\begin{equation}
\Delta \mu \ = \ j {\pi
\hbar \over 4 e} \int_0^{L/\ell_0} d \overline{x} \tanh^2\left(1 \over
t(\overline{x})\right) \ \ .
\label{currentvoltage.eq}
\end{equation}
%*******************************************************************
%
Since the functional form of the rescaled temperature profile $t(\overline{x})$
only depends on the dimensionless parameter $L/\ell_0$ through the boundary
conditions of Eq.~(\ref{Tdiff.eq}), it is clear from
Eq.~(\ref{currentvoltage.eq}) that the resistance of a long edge is Ohmic in
the limit of zero bath temperature.  However, we obtain an interesting scaling 
$\pi \hbar G/e^2 \propto (l_0/L)^{0.35}$ of the conductance with the length of the wire. 
The current noise can be obtained from
\cite{Steinbach:1996vq,vonOppen:1997vc}
%
%*****************************  current noise  **************************
\begin{equation}
S \ = \ 4 {1 \over L^2} \int_0^L dx  \, \sigma(x) \,  k_\text{B} T(x)  \ \ ,
\label{eq:currentNoise}
\end{equation}
%******************************************************************
%
and can be expressed in the form $S =  F \,  G \, V$, where $F$ denotes the Fano
factor and $G$ is the conductance obtained from the current voltage relation
Eq.~(\ref{currentvoltage.eq}).  By studying the behavior of
Eq.~(\ref{eq:currentNoise}) in the long edge limit, as shown in
Fig.~\ref{fig:fano}, we extract a numerical result for the Fano factor of $F =
1.219(2)$ where the number in the bracket indicates the uncertainty in the
final digit.  By comparison, the Fano factor for a one dimensional diffusive
wire much longer than the inelastic mean free path is $\sqrt{3}/2$
\cite{Nagaev95,KoRu95}, if cooling due to electron-phonon coupling can be
neglected.

% ===============================================================================
% Discussion and Estimate of the Mean Free Path
% ===============================================================================

Before attempting to apply the results discussed here to the HgTe quantum wells
studied in Ref.~\cite{Konig:2007hs} it is useful to put them in the context of
other potential backscattering mechanisms that have been proposed to explain
the observed deviation from ballistic edge transport. In the presence of both
time reversal symmetry breaking and disorder, the helical edge states will be
localized, as studied in the presence of magnetic fields and impurities
\cite{MaciejkoPRB10}, random magnetic fluxes \cite{Delplace:2012ww} and
magnetic impurities \cite{Tanaka11, Hattori}. Alternatively, back-scattering
becomes possible in the presence of dephasing processes, which have been
simulated phenomenologically using B\"uttiker's virtual probes
\cite{Roth:2009bg, Jiang:2009fr}. Budich \emph{et al.} \cite{Budich:2012dg}
found that inelastic processes due to electron-phonon interactions in the
presence of spin orbit coupling cannot cause backscattering between helical
edge states to leading order.   Other studies have shown that avenues towards
backscattering from magnetic impurities alone \cite{Maciejko:2009kw},
electron-electron interactions \cite{Crepin:2012vj, Lezmy12} or the loss of
axial spin symmetry \cite{Schmidt:2012iw} are all suppressed in the low
temperature limit.
%
% ----------------------------------------------------------------------------
% ----------------------------------------------------------------------------
%

Backscattering from dynamic nuclear polarization persists to zero temperature.
In order to obtain an estimate for the mean free path $\ell_0$, we exploit the
interpretation of $A \langle I \rangle$ as a local effective Zeeman splitting.
In this way, we can  determine the parameters in the scattering rate by comparison with
the effects of backscattering due to an in-plane magnetic field in the presence
of a spatially random Rashba coupling.  Examining Ref.~\cite{Konig:2007hs} we
estimate that $(\hbar v)^4/2V_R \approx 0.14~ (\mu \text{eV})^2 \mu \text{m}$, which in
combination with a hyperfine constant of $A \approx 60~\mu\text{eV}$
for interactions of electrons with the fully polarized nuclei of mercury
\cite{Bogle:2002hb} produces $\ell_0 \approx 40~\mu\text{m}$.

% ===============================================================================
% Acknowledgments
% ===============================================================================
A.D. acknowledges financial support from the University of Vermont and the
gracious hospitality of the Institut f\"{u}r Theoretische Physik,
Universit\"{a}t Leipzig where the majority of this work was completed.

% ===============================================================================
% Bibliography
% ===============================================================================
\bibliographystyle{apsrev4-1.bst}
\bibliography{edge_polarization}

\end{document}